\documentclass[sigplan,10pt]{acmart}

\def\sysname{NecoFuzz\xspace}
\def\titlename{\sysname: Effective Fuzzing of Nested Virtualization via Fuzz-Harness Virtual Machines}

\def\keyword{Nested Virtualization, Fuzzing}

\usepackage{shina-lab}
\usepackage{graphicx}
\usepackage{enumitem}
\usepackage{xspace}
\usepackage{comment}
\usepackage{caption}
\usepackage{subcaption}

\begin{document}
\title{\titlename}
\NOACMmaketitle

\begin{abstract}
Nested virtualization is now widely supported by major cloud vendors, allowing users to leverage virtualization-based technologies in the cloud.  
However, supporting nested virtualization significantly increases host hypervisor complexity and introduces a new attack surface in cloud platforms.  
While many prior studies have explored hypervisor fuzzing, none has explicitly addressed nested virtualization due to the challenge of generating effective virtual machine (VM) instances with a vast state space as fuzzing inputs.

We present \sysname, the first fuzzing framework that systematically targets nested virtualization-specific logic in hypervisors.  
\sysname synthesizes executable \emph{fuzz-harness VMs} with internal states near the boundary between valid and invalid, guided by an approximate model of hardware-assisted virtualization specifications.  
Since vulnerabilities in nested virtualization often stem from incorrect handling of unexpected VM states, this specification-guided, boundary-oriented generation significantly improves coverage of security-critical code across different hypervisors.

We implemented \sysname on Intel VT-x and AMD-V by extending AFL++ to support fuzz-harness VMs.  
\sysname achieved 84.7\% and 74.2\% code coverage for nested virtualization-specific code on Intel VT-x and AMD-V, respectively, and uncovered six previously unknown vulnerabilities across three hypervisors, including two assigned CVEs.
\end{abstract}

\ACMmaketitle
\IEEEkeyword

\section{Introduction}

Nested virtualization has become an important feature in modern cloud computing.  
While traditional virtualization enables the essential properties of cloud platforms, nested virtualization unlocks virtualization-based technologies for cloud users, enabling a wide range of use cases such as enhanced security, legacy system migration, and disaster recovery~\cite{microsoft-vbs,windows-sandbox,crosvm,katacontainers,scale-disaster-recovery}.  
Since its initial support in 2017, nested virtualization has been adopted by several major cloud vendors~\cite{nested-azure,nested-gce,nested-oracle,nested-alibaba}, and it now serves as a core building block for confidential virtual machines (VMs) based on Intel TDX in Microsoft Azure~\cite{paravisor}.  
As a result, it has matured into a commercially available feature in public cloud services.

Unfortunately, nested virtualization introduces a new attack surface in cloud platforms.  
To support nested virtualization, host hypervisors must expose a hardware-assisted virtualization interface, such as Intel VT-x~\cite{intel-sdm} or AMD-V~\cite{amd-manual}, to guest hypervisors running inside user VMs.  
Since current CPUs do not natively support nested virtualization, host hypervisors must re-virtualize the interface in software~\cite{183330}.  
Hardware-assisted virtualization is already known to be complex and error-prone in non-nested settings~\cite{ASVIJA201968}.  
Re-virtualizing it further increases the implementation burden, requiring precise handling of intricate behaviors and subtle corner cases.  
Even with enhanced hardware support for nested virtualization~\cite{10.1145/3373376.3378467,10.1145/3307650.3322261}, securely handling the interface remains difficult in practice.  
In fact, multiple vulnerabilities have been discovered in nested virtualization code over the past several years~\cite{CVE-2017-12188,CVE-2017-2596,CVE-2018-16882,CVE-2019-3887,CVE-2021-3656,CVE-2021-29657,CVE-2022-45869}.  
Given its deployment in production cloud platforms, such vulnerabilities pose serious risks to cloud infrastructure, including potential compromise of host hypervisors and co-located tenants.

Fuzzing is an effective technique for discovering security vulnerabilities, and many studies have explored its application to hypervisors.  
However, none of them explicitly addresses nested virtualization, leaving the hardware-assisted virtualization interface completely untested.  
For example, most prior hypervisor fuzzing targets I/O interfaces such as programmable I/O (PIO), memory-mapped I/O (MMIO), or DMA~\cite{10.1007/978-3-319-66332-6_1, hyper-cube, 263866, 10.1145/3460120.3484811, 277148, 10179354, 298158}.  
These I/O-based approaches do not access the nested virtualization interface at all, which is accessed through dedicated CPU instructions.  
Other studies have explored fuzzing via instruction streams, targeting CPU emulators and their instruction-handling logic~\cite{10.1145/1572272.1572303,10.1145/1831708.1831730,10.1145/2815400.2815420,10.1145/2150976.2151012,10.1145/3190508.3190529,10.1145/3460120.3484748,10202630}.  
However, these also fail to cover hardware-assisted virtualization instructions, which operate on structured VM states and require hypervisor-level coordination.  
Device driver fuzzing~\cite{180270,10179293,281316} exercises OS kernel modules, which operate at a different layer and interface than the hypervisor.

The key challenge in nested virtualization fuzzing lies in its input.  
Since nested virtualization involves running a VM inside another VM, the input is a complete VM instance rather than simple values.  
Hardware-assisted virtualization interfaces typically take a VM state as an argument to special CPU instructions when executing a VM instance.  
A VM state consists of several KB of data representing the internal state of the virtual CPU (vCPU), forming a vast state space divided into valid and invalid states, where the boundary is determined by complex constraints across many interdependent fields.  
This enormous state space, combined with these constraints, makes effective VM state generation through random values highly impractical.  
Even when starting from a valid \emph{golden} seeds~\cite{281316}, there are numerous locations and directions in the VM state to mutate, and determining effective mutation strategies is nontrivial.
Symbolic and concolic testing also face difficulties with this vast state space, even though they appear effective for small nested virtualization code bases.

In this paper, we present \sysname, the first fuzzing framework for nested virtualization.  
\sysname generates complete VM instances, called \emph{fuzz-harness VMs}, whose internal states are diverse yet near the boundary between valid and invalid configurations to effectively trigger error-prone hypervisor behavior.
It uses \emph{VM state validators} that model hardware-assisted virtualization specifications to efficiently explore this boundary.  
To handle the complexity and undocumented aspects of the specification, \sysname verifies generated states on physical CPUs, using hardware behavior as ground truth to detect and correct modeling inaccuracies at runtime.  
To broaden coverage, \sysname mutates vCPU configurations, controlled via user interfaces.  
To bridge this gap, it employs a \emph{vCPU configurator} on the host side.  
\sysname detects anomalies via sanitizers and log monitoring, targeting nested virtualization code poorly tested by prior work.
By combining these features, it enables practical fuzzing of this previously unexplored attack surface.

We implemented \sysname on Intel VT-x and AMD-V.  
The fuzz-harness VM is structured as a single UEFI binary that boots on the host hypervisor, initializes hardware-assisted virtualization as a guest hypervisor, and launches a nested guest VM.  
The validator is built using CPU emulator code extracted from Bochs~\cite{bochs}, and the vCPU configurator is implemented as a script that sets vCPU options via command-line parameters.  
Since existing hypervisor fuzzers cannot accommodate full VM instances, we extend AFL++~\cite{afl++} to interact with the fuzz-harness VM to explore nested virtualization code paths in the host hypervisor.

We evaluate \sysname on KVM (Linux 6.5) to measure code coverage of nested virtualization-specific code.  
As baselines, we use Syzkaller~\cite{syzkaller} (commit 96a211b), the only available fuzzer supporting nested virtualization, and IRIS~\cite{10202630}, a state-of-the-art fuzzer for the Xen hypervisor, which does not explicitly support nested virtualization and is limited to Intel processors.  
\sysname achieved 84.7\% coverage on Intel VT-x and 74.2\% on AMD-V, representing improvements of 1.4$\times$ and 11.0$\times$ over Syzkaller on Intel VT-x and AMD-V, respectively, and 1.6$\times$ over IRIS on Intel VT-x.
This improvement was achieved by automatically reaching specification-boundary states.
We also evaluate \sysname on Xen and VirtualBox to assess its effectiveness in vulnerability discovery.  
It uncovered six previously unknown vulnerabilities across the three hypervisors, all confirmed by developers.  
Four have been fixed, and two assigned CVEs~\cite{CVE-2023-30456,CVE-2024-21106}.  
Additionally, we identified and patched two bugs in Bochs~\cite{patch-bochs} during validator development.

The contributions of this paper are as follows:

\begin{itemize}
  \item We present \sysname, the first fuzzing framework that systematically targets nested virtualization logic in hypervisors by generating complete VM instances as inputs to the hardware-assisted virtualization interface, an attack surface unaddressed by prior fuzzers.
  
  \item We design a hypervisor-agnostic VM generator tailored to the structural complexity of nested virtualization, synthesizing fuzz-harness VMs near the valid/invalid boundary guided by specifications, and exploring diverse vCPU configurations.
  
  \item We implement \sysname for Intel VT-x and AMD-V, extending AFL++ to enable efficient fuzzing via bootable VMs without relying on hypervisor source code or syscall interfaces.
  
  \item We evaluate \sysname on KVM, Xen, and VirtualBox, showing substantial coverage improvement for nested virtualization logic over existing tools (e.g., Syzkaller, IRIS) and discovering six previously unknown vulnerabilities, including two CVEs.
\end{itemize}

\section{Background}

This section summarizes background knowledge on hardware-assisted virtualization and nested virtualization.

\subsection{Hardware-Assisted Virtualization}
\label{sec:hw-v}

In virtualizing a CPU to create a VM, a statistically dominant subset of processor instructions should execute directly on physical hardware~\cite{10.1145/361011.361073}.  
To enable this, \emph{sensitive} instructions must be trappable by the hypervisor.  
However, the classic x86 instruction set was not fully virtualizable, as some sensitive instructions were not privileged~\cite{271274}.  
To address this issue, Intel and AMD introduced hardware-assisted virtualization, known as Intel VT-x and AMD-V, respectively.  
As the two designs are conceptually similar, we use Intel terminology throughout unless otherwise noted.

Intel VT-x introduces two CPU modes orthogonal to existing ones: \emph{non-root mode} and \emph{root mode}.  
When switching between these modes, the CPU stores its internal state in a memory-resident structure called the \emph{VM control structure (VMCS)}.  
A hypervisor launches a VM by executing the \texttt{vmlaunch} instruction with the corresponding VMCS as input.  
The instruction sequence for initializing hardware-assisted virtualization is largely fixed: the hypervisor initializes a VMCS using \texttt{vmclear}, sets the VMCS pointer with \texttt{vmptrld}, writes VMCS values through a series of \texttt{vmwrite} instructions, and starts the VM with \texttt{vmlaunch}, referred to as \emph{VM entry}.  
When control returns to the hypervisor from the VM, known as a \emph{VM exit}, the hypervisor uses \texttt{vmread} to determine the exit reason, updates the VM state with \texttt{vmwrite}, and resumes VM execution using \texttt{vmresume}.  
Failure to adhere to this sequence prevents the VM from reaching an initialized state.

The conditions for a valid VM state are highly complex.  
For instance, the VMCS contains control fields with reserved bits that must be set to fixed values; any incorrect setting results in a VM entry failure.  
The VMCS also stores saved register values, including internal ones, each subject to detailed constraints, both individually and in combination, to ensure correctness and security.  
These constraints are enforced by the physical CPU, and a VM entry fails if even one is violated.

Each new CPU generation brings enhancements to hardware-assisted virtualization.  
For example, early Intel CPUs lacked support for nested paging, which was later introduced as extended page tables (EPT).  
Real-mode execution was initially unsupported, but became available with the unrestricted guest feature.  
The availability of such features is determined by model-specific registers (MSRs).

\subsection{Nested Virtualization}

\begin{figure}[t]
  \centering
  \includegraphics[width=\columnwidth]{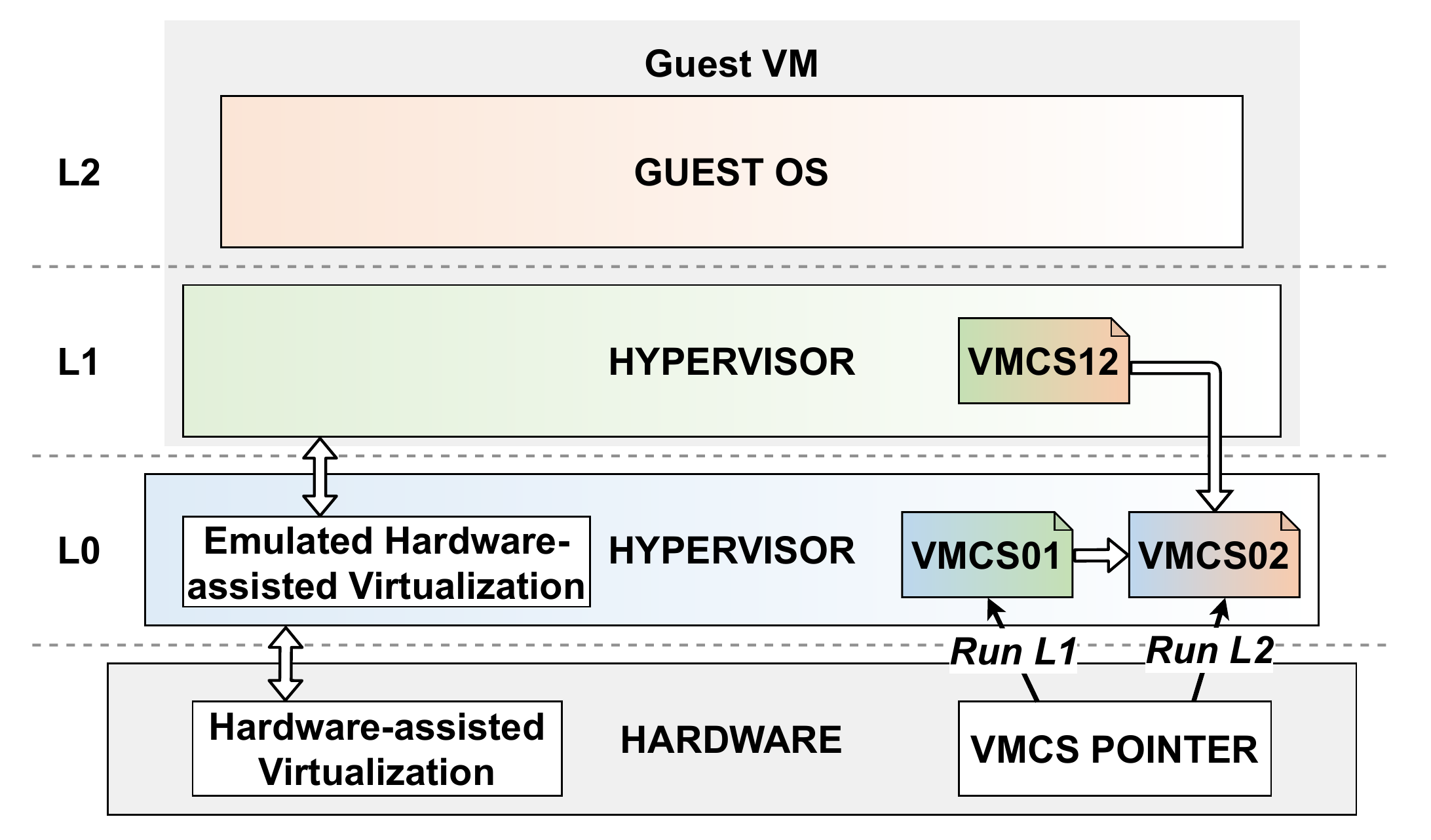}
  \caption{Nested virtualization. The L0 hypervisor uses hardware-assisted virtualization and emulates it for the L1 hypervisor.
  L0 uses two VMCSs: VMCS01 for L1 and VMCS02 for L2, while L1 maintains VMCS12 for L2.}
  \label{fig:nested}
\end{figure}

Nested virtualization allows a guest hypervisor to run on top of another host hypervisor.  
The lower-level hypervisor is referred to as L0, and the upper one as L1 (see \autoref{fig:nested}).  
The guest OS or VM under the L1 hypervisor is called L2.  
The L0 hypervisor directly uses hardware-assisted virtualization provided by the physical CPU and emulates it for the L1 hypervisor.  
To achieve this, L0 runs L1 in non-root mode and emulates hardware behavior in software.

Emulating VMCSs is particularly complex.  
The L1 hypervisor assumes it interacts with hardware directly and maintains a VMCS for each of its L2 guests, referred to as \emph{VMCS12}.  
However, VMCS12 is not recognized by the hardware and must be emulated by L0.  
To run L1, L0 uses \emph{VMCS01}, where L0 is the host and L1 the guest.  
Similarly, it uses \emph{VMCS02} to run L2, where L0 is again the host and L2 the guest.  
L0 switches between VMCS01 and VMCS02 to emulate mode switching between L1 and L2.

To correctly emulate VMCSs, L0 must synchronize their contents by interpreting VMCS fields and translating them across VMCS12, VMCS01, and VMCS02.  
It must also replicate the security and consistency checks performed by the CPU.  
For instance, L0 must prevent L1 from configuring VMCS12 to access L0’s memory region.  
Furthermore, the internal emulation state maintained by L0 must remain consistent with the actual hardware VMCS state.  
For example, the CPU may silently correct invalid field values, and L0 must emulate this behavior precisely.

In addition, the hardware-assisted virtualization features available to the L1 VM depend on the vCPU configuration.  
While L0 emulates certain features in software, software and hardware states are interdependent.  
Thus, when a feature is disabled in the vCPU configuration, L0 must ensure consistent behavior with the underlying hardware.


\section{Design}

\begin{figure}[t]
  \centering
  \includegraphics[width=\columnwidth]{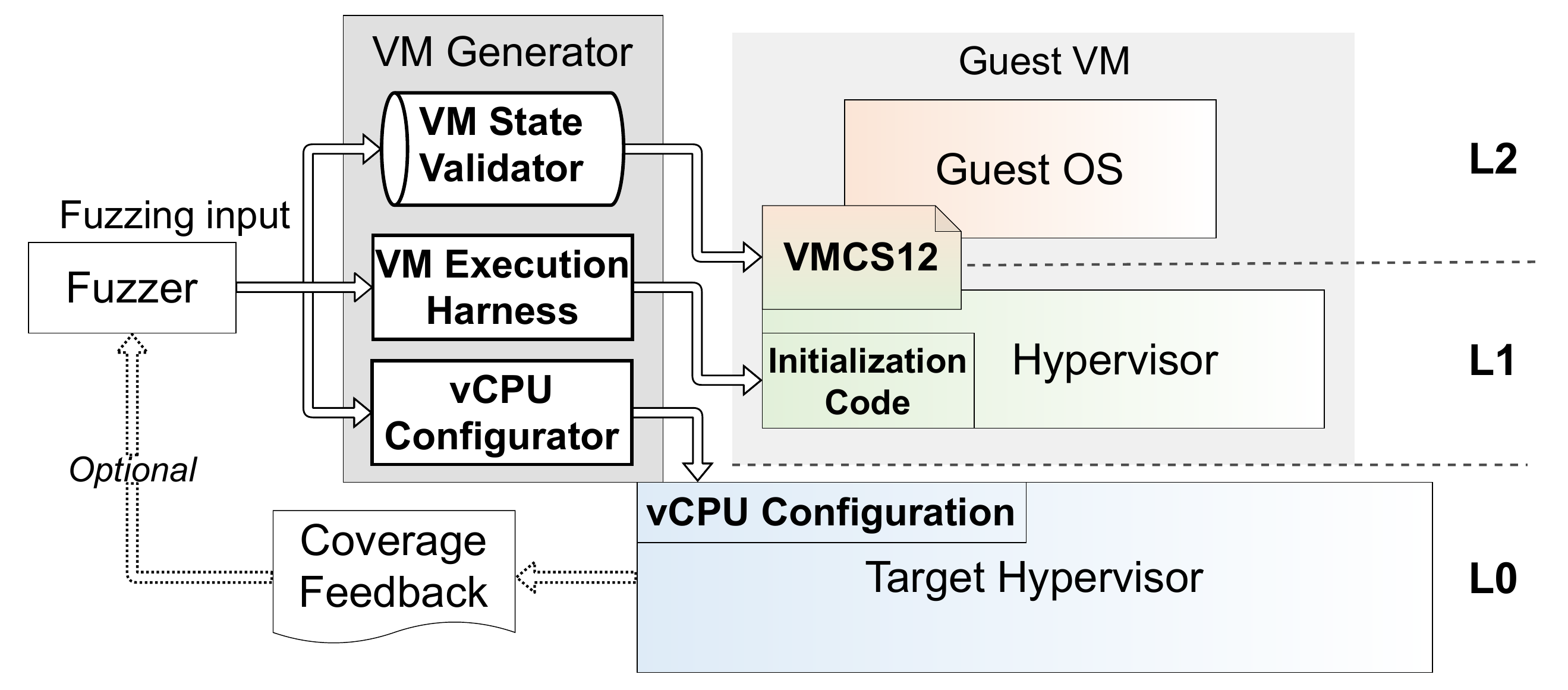}
  \caption{A design overview of \sysname. The VM generator consists of three components: (1) VM execution harness, (2) VM state validator, and (3) vCPU configurator. The fuzzer supplies \emph{fuzzing input} to each component.}
  \label{fig:system}
  \centering
\end{figure}

This section describes the design of \sysname.  
We begin with the threat model, then provide an overview of \sysname, followed by the explanations of its three main components.

\subsection{Threat Model}
\label{subsec:threat_model}

We assume a cloud environment where the provider manages the L0 hypervisor and the attacker obtains a VM instance.  
The attacker has full control over the software environment within the VM, including the ability to execute arbitrary code.
Nested virtualization is enabled in the VM, allowing the attacker to issue hardware-assisted virtualization instructions from inside the instance.  
They may run custom programs as the L1 hypervisor and L2 guest OS, unconstrained by existing hypervisor or OS implementations.  
The attacker can construct instruction sequences and VM states that would not be generated by standard systems.
Our focus is on the L0 hypervisor, as it underpins cloud infrastructure and is deployed in production.  
We do not target vulnerabilities in software running within VMs, regardless of nested virtualization use cases, as these VMs are fully controlled by the attacker.

We specifically target attacks that exploit hardware-assisted virtualization interfaces exposed to guest VMs.  
We exclude interfaces that are less directly related to nested virtualization, such as virtual device interfaces (PIO, MMIO, DMA) and hypercalls.  
We also exclude interfaces that can only be invoked from the host side, such as ioctl()s for live migration, since both the L0 hypervisor and host OS are assumed to be fully trusted by the cloud provider.
Side channels, covert channels, and other indirect attack vectors are also out of scope.

\subsection{Overview}

\sysname introduces a VM generator that efficiently produces fuzz-harness VMs, guided by the hardware-assisted virtualization specification.  
It consists of three components: (1) a VM execution harness, (2) a VM state validator, and (3) a vCPU configurator.
\autoref{fig:system} provides its overview.

\sysname leverages an existing fuzzer as its base, adapting it to generate binary \emph{fuzzing input}.  
This input is partitioned and dispatched to each component of the VM generator to mutate specific parts of the fuzz-harness VM.  
For example, the VM execution harness mutates the execution order and parameters within the VM.  
The VM state validator mutates the VMCS to produce diverse VM states.  
The vCPU configurator mutates combinations of supported vCPU features.


We run the L0 hypervisor on bare metal, since executing it in a virtual environment would require \emph{double nested} virtualization (i.e., ``triple virtualization''), which remains too slow and unstable in practice.  
When a bug or vulnerability in the L0 hypervisor is triggered during fuzzing, the system may crash or produce an error message.  
To handle such cases and maintain continuous fuzzing, \sysname includes a watchdog mechanism that combines a hardware feature with an agent inside the L0 hypervisor to detect failures and automatically restart the hypervisor.  
Since crashes are rare, the overhead of restarting has minimal impact on fuzzing efficiency.

\subsection{VM Execution Harness}
\label{subsec:design:harness}

The VM execution harness forms the basis of the fuzz-harness VM, which plays the roles of both the L1 hypervisor and the L2 guest OS.  
It enables the VM to be initialized and executed on the L0 hypervisor, efficiently exercising the hardware-assisted virtualization interface.  
It operates in two phases: an \emph{initialization phase} and a \emph{runtime phase}.

In the initialization phase, the harness executes a sequence of hardware-assisted virtualization instructions as an L1 hypervisor to initialize the L2 VM.  
As explained in \autoref{sec:hw-v}, this initialization sequence is largely fixed, and any significant deviation is promptly rejected by the L0 hypervisor’s error-checking logic, which does not contribute to improving code coverage. 
However, strictly adhering to the fixed sequence risks missing vulnerabilities in the initialization logic of nested virtualization emulation.

To address this, we prepare a manually written template of the instruction sequence and mutate it using the fuzzing input.  
This template consists of only a few hundred lines of code, requiring minimal effort.  
By mutating the arguments and order of instructions, the initialization sequence emulation code in the L0 hypervisor is effectively fuzzed.  
Once sufficient coverage of this code is achieved, the harness transitions to the next runtime phase.

In the runtime phase, the harness executes CPU instructions that trigger VM exits to the L0 hypervisor.  
A VM exit is typically caused by a single instruction, which may require a few setup instructions and is executed with either L1 or L2 privileges.  
To support this, we prepare templates for all such instructions.  
The harness mutates the selection and arguments of these templates using the fuzzing input and repeatedly executes them.  
This approach efficiently triggers VM exits and exercises nested virtualization code in the L0 hypervisor.

\subsection{VM State Validator}

The VM state validator checks the validity of VM states within the fuzz-harness VM used at VM entry.  
As described in \autoref{sec:hw-v}, the VMCS is subject to complex validity conditions.  
In nested virtualization, the L0 hypervisor must replicate this validation logic to determine whether a VMCS provided by an L1 hypervisor is valid.  
If validation is incomplete or inaccurate, invalid VMCS12 may propagate into VMCS01 or VMCS02, potentially compromising L0 hypervisor security.  
Thoroughly testing the VM state validation code in nested virtualization is therefore essential.

However, a randomly generated VMCS is highly likely to contain obvious errors.  
Such invalid fields are detected early by the L0 hypervisor, which returns an error without improving code coverage.  
Moreover, since the VMCS spans several KBs, exhaustively exploring all possible values is highly inefficient.  
To address this, the VM state validator first generates a VMCS based on the fuzzing input, rounds it to a valid state, and then selectively mutates parts of it to introduce invalid values.  
This approach efficiently produces VMCSs that are diverse yet neither fully valid nor trivially invalid, enabling targeted testing near the boundary between valid and invalid states.  
Compared to gradually mutating a golden seed, it explores a broader range of values while consistently remaining near the boundary region over time.

One challenge is implementing the VM state validator itself correctly.  
As previously noted, the conditions under which a VMCS is considered valid are highly complex; the Intel manual specifies them over many pages of natural-language descriptions.  
Moreover, Intel processors do not report which specific field in the VMCS caused a VM entry failure, making correctness verification difficult.  
Some constraints are also undocumented, and in certain cases, the CPU silently rounds VMCS values to correct inconsistencies.  
Therefore, implementing a fully accurate VMCS validator solely based on documentation is impractical.

To address this, the validator sets the generated VMCS on the actual CPU, attempts a VM entry, and compares the resulting VMCS state with the expected one.  
By using the physical CPU as an oracle, this approach not only checks the correctness of the VMCS but also validates the implementation of the VM state validator itself.  
This idea is similar to existing CPU emulator fuzzing approaches~\cite{10.1145/1572272.1572303,10.1145/2150976.2151012}, but applied in reverse: rather than verifying the emulator being fuzzed, we verify a component of the fuzzer itself.  
This method enables more accurate testing of boundary checks in nested virtualization code by detecting and correcting errors in the validator's implementation.

We need to prepare only a single VM state validator for each CPU architecture, as the VMCS validity conditions are defined by the CPU architecture.  
Therefore, the VM state validator is essentially hypervisor-independent and can be applied to any L0 hypervisor.  

\subsection{vCPU Configurator}

vCPU configuration, such as enabling or disabling specific CPU features, is typically specified at VM startup via parameters passed to the hypervisor.  
Configurable options related to hardware-assisted virtualization include support for EPT, unrestricted guest, virtual processor ID, and others.  
As the number of features grows, the space of possible enable/disable combinations increases exponentially.  
Moreover, these features interact with one another and with the VM state, further increasing complexity.  
Since VM configuration significantly influences hypervisor behavior, unexpected combinations can expose vulnerabilities.  
However, conventional hypervisor fuzzing typically uses static vCPU configurations, potentially missing vulnerabilities tied to specific settings.

To address this, \sysname employs a vCPU configurator to explore a wide range of configurations.  
Based on fuzzing inputs, it mutates startup parameters passed to the L0 hypervisor to generate diverse VM configurations.  
Combined with the VM state validator, this approach covers a broader range of scenarios, exercises interactions across hypervisor components, and increases the likelihood of uncovering vulnerabilities.

While the vCPU configuration itself is CPU-independent, it must be applied by the host L0 hypervisor through a hypervisor-specific interface.  
To bridge this gap, the vCPU configurator consists of a hypervisor-independent core that generates configurations from fuzzing inputs, and a small adapter connecting to each L0 hypervisor.  
This design enables easy adaptation across hypervisors with minimal effort.

\section{Implementation}

This section describes the implementation of \sysname.  
We first present an overview of the fuzzing framework, followed by the implementation details of the VM execution harness, VM state validator, and vCPU configurator.  
We then explain the agent program and its integration with the hypervisor.  

\subsection{Fuzzing Framework}
\label{subsec:fuzzing_framework}


We leverage AFL++~\cite{afl++} as the base fuzzing framework.  
AFL++ is designed for fuzzing user-space applications, so we implement an agent program on the host OS that connects the fuzzer, the fuzz-harness VM, and the target L0 hypervisor.  
This agent receives fuzzing inputs from AFL++ (2KiB of binary data) and passes them to the fuzz-harness VM, while also collecting coverage data from the L0 hypervisor and reporting it back to AFL++.

To collect coverage information, we use hypervisor-specific mechanisms but abstracts them into a unified interface for AFL++.  
On KVM, it employs the kcov interface~\cite{kcov} to obtain instruction-level traces; on Xen, it uses gcov~\cite{gcov} through a custom wrapper that converts its output into a compatible format.  
The agent maps these traces to a shared memory bitmap monitored by AFL++ to guide mutation.  
This design isolates fuzzing logic from hypervisor internals, enabling \sysname to operate in a largely hypervisor-agnostic manner.



The core fuzzing logic within the fuzz-harness VM is orchestrated by an \emph{executor}, implemented as a self-contained UEFI application that integrates the VM execution harness (\autoref{subsec:vm_execution_harness}) and the VM state validator (\autoref{subsec:vm_state_validator}).  
It runs with privileges to execute instructions in both L1 (hypervisor) and L2 (guest) contexts.  
For Intel VT-x, the executor consists of approximately 3,500 lines of C code, organized as follows:

\begin{itemize}[leftmargin=*,nosep]
    \item \textbf{Fuzzing Orchestration} (\textasciitilde600 LOC): Coordinates the overall fuzzing process, including initialization, VMEXIT dispatch, and termination.
    \item \textbf{VM Execution Harness} (\textasciitilde900 LOC): Executes individual CPU instructions (e.g., VMX operations, register access, I/O, MSR R/W) executable in L1 or L2.
    \item \textbf{VM State Validator} (\textasciitilde2,000 LOC): Ensures VMCS validity. It extends Bochs-derived logic with field correction, value rounding, and integration mechanisms.
\end{itemize}

\subsection{VM Execution Harness}
\label{subsec:vm_execution_harness}

\begin{table*}[t]
\centering
\begin{tabular}{l|l|l}
\hline
\textbf{Class} & \textbf{Example Instructions} & \textbf{Handling} \\
\hline
VMX Instructions & \texttt{vmxon}, \texttt{vmclear}, \texttt{vmlaunch}, \texttt{vmresume}, etc. & Emulated by the L0 hypervisor \\
Privileged Registers & \texttt{mov cr0/cr3/cr4}, \texttt{mov dr0..dr7} & Commonly intercepted \\
I/O and MSR Operations & \texttt{in/out}, \texttt{rdmsr}, \texttt{wrmsr} & Selectively intercepted based on bitmaps \\
Miscellaneous & \texttt{cpuid}, \texttt{hlt}, \texttt{rdtsc}, \texttt{pause}, \texttt{rdrand}, etc. & Commonly intercepted \\
\hline
\end{tabular}
\caption{Examples of instructions that cause VM exits on Intel VT-x.}
\label{tab:vm_instructions}
\end{table*}

The VM execution harness is divided into an initialization phase and an execution phase.  
Both phases execute hardware-assisted virtualization CPU instructions based on fuzzing inputs, which are received from the agent as binary data.

\paragraph{VM Initialization Phase.}




The initialization sequence is critical but brittle: deviating from it causes immediate VM entry failures.
To navigate this, we designed a domain-specific template of initialization instructions that reflects the standard VMX setup sequence (e.g., \texttt{vmxon}, \texttt{vmclear}, \texttt{vmptrld}, \texttt{vmwrite}, \texttt{vmlaunch}).
This template is interpreted by a lightweight custom engine, and fuzzing input mutates its instruction ordering, argument values, and repetition counts.
This design allows exploration of subtle control flow variations while preserving structural correctness, avoiding invalid states that would terminate fuzzing prematurely.

\paragraph{VM Execution Phase.}
This phase repeatedly executes VMs, putting stress on the L0 hypervisor.  
The harness runs a tight loop that performs the following steps:
\begin{enumerate}
\item Selects and executes an instruction sequence in the L2 guest context, based on fuzzing input, which may trigger a VM exit to L1.
\item If a VM exit to L1 occurs, selects and executes another instruction sequence in the L1 hypervisor context, which is emulated by the L0 hypervisor, based on fuzzing input.
\item Re-enters the L2 guest using \texttt{vmresume}.
\end{enumerate}


Rather than executing unstructured instruction streams, the execution harness uses a library of templates representing known exit-triggering instructions (e.g., \texttt{mov cr*}, \texttt{rdmsr}, \texttt{in/out}) wrapped with minimal setup logic.
This template-based strategy improves signal-to-noise ratio by increasing the likelihood that executed instructions exercise meaningful VM exit paths.
Parameters for each instruction (e.g., target registers, operand values) are derived from fuzzing input, enabling scalable yet semantically constrained instruction-level fuzzing in both L1 and L2 contexts.
Although the instruction details differ on AMD-V, the overall approach remains the same.
This cycle continues until a termination condition is met, either due to a critical failure or upon reaching an iteration limit.

\subsection{VM State Validator}
\label{subsec:vm_state_validator}

The VMCS is a hardware-defined structure containing over 150 fields grouped into areas such as control fields, guest and host state, and entry and exit controls.
Each field is subject to strict constraints and dependencies; for example, some control fields must match MSR capabilities, and guest CR0/CR4 values must satisfy architectural rules.
This structure spans several kilobytes and presents significant challenges for efficient mutation.
Rather than reimplementing these constraint checkers from scratch, we extracted and extended Bochs’s high-fidelity VMCS validation logic, adapting it to work dynamically within the fuzz-harness VM.
This approach not only avoids reliance on static golden states but also enables boundary-aware fuzzing: generated states are valid enough to pass L0 checks but close to rejection, maximizing coverage of validation logic.

The validation logic we adapted consists of about 2,500 lines of C code.
We decoupled this logic from Bochs’s internal structures and integrated it into our fuzz-harness VM.
Specifically, we incorporated the following routines, each responsible for validating a VMCS field group during Bochs’s simulated VM entry:

\texttt{\textbf{VMenterLoadCheckVmControls()}}:  
Validates VM-execution control fields (Pin-Based, Processor-Based Primary and Secondary), the exception bitmap, CR0/CR4 guest/host masks and read shadows, and associated addresses (e.g., I/O bitmap addresses, MSR bitmap address).

\texttt{\textbf{VMenterLoadCheckHostState()}}:  
Validates host-state area fields, including control registers (CR0, CR3, CR4), segment selectors and bases, GDT/IDT bases, and MSRs such as \texttt{IA32\_SYSENTER\_CS}, \texttt{ESP}, and \texttt{EIP}.

\texttt{\textbf{VMenterLoadCheckGuestState()}}:  
Validates guest-state area fields, including RFLAGS, control registers (CR0, CR3, CR4), segment registers, GDT/IDT/LDT/TR, MSRs, activity state, and interruptibility state.

To generate VMCS data, we first extract a few kilobytes of binary data from the fuzzing input provided by the agent, treating it as raw VMCS content.  
We then apply the validation logic above to adjust invalid fields by rounding them to the nearest valid values.  
Next, we mutate the corrected VMCS fields using additional bytes from the fuzzing input.  
Finally, the resulting VMCS is written using \texttt{vmwrite}.

The mutation logic proceeds as follows:

\begin{enumerate}
    \item \textbf{Field Selection:} A VMCS field is selected for mutation, guided by fuzzing input to explore different regions of the VMCS state space.
    \item \textbf{Bit Selection:} Within the selected field, one or more bit positions are chosen based on fuzzing input. The selection is constrained to the field's valid bit-width.
    \item \textbf{Mutation:} The chosen bit(s) are flipped.
    \item \textbf{Iteration:} Steps 2 and 3 are typically repeated across one to three VMCS fields per fuzzing iteration, mutating one to eight bits per field. The number of fields and bits is determined by fuzzing input.
\end{enumerate}

This mutation strategy deliberately perturbs VMCS fields after they have been mostly validated.  
By focusing bit flips on security-critical areas such as control fields and access rights registers, the fuzzer generates VM states near the boundary between valid and invalid configurations.
Such states are more likely to expose subtle flaws in error handling or unexpected interactions within the hypervisor’s VMX management logic.

To address cross-field constraints, we designed the validator’s rounding procedure to operate sequentially across three VMCS field groups, in the order of control fields, host-state fields, and guest-state fields.  
In each group, fields are first rounded to specification-compliant values using Bochs’ validation logic as described above, and intra-group constraints are then checked and corrected if necessary.  
For instance, if \texttt{IA32\_EFER.LME} (long mode enable) is set to run in x86-64 mode while \texttt{CR4.PAE} (physical address extension) is unset in the guest-state or host-state fields, the validator forces this bit to 1 to satisfy architectural constraints.  
Inter-group constraints are also checked and corrected against the previously processed groups.  
Independent fields can be modified individually according to the specification, and even dependent fields form a unidirectional graph, allowing deterministic correction in each step.  
Through these sequential validations, the rounding procedure completes in a bounded number of steps and ensures consistency across VMCS fields with complex interdependencies.




\subsection{vCPU Configurator}

The vCPU configuration is generally represented as a bit array, where each bit indicates whether a specific CPU feature is enabled or disabled.  
This configuration is mutated based on fuzzing input.  
To apply the generated configuration to the L0 hypervisor, we implement a small adapter for each hypervisor.

For KVM, the adapter applies vCPU configurations through two interfaces:

\paragraph{Kernel Module Parameters:}  
The vendor-specific parts of KVM (Intel VT-x and AMD-V) are implemented as separate kernel modules (\texttt{kvm-intel.ko}, \texttt{kvm-amd.ko}).  
These modules accept parameters that configure hardware-assisted virtualization features, such as enabling or disabling EPT.

\paragraph{Hypervisor Command-line Options:}  
More general vCPU configurations are specified as command-line options to the hypervisor, QEMU, in this case.  
These configurations include CPU model features (e.g., enabling or disabling VMX/SVM extensions, or specific CPUID flags such as \texttt{hv-passthrough}), virtual CPU topology, and memory allocation parameters.

We implement the KVM adapter as a shell script that reloads the kernel module with the desired parameter string and launches QEMU with the appropriate command-line options.  
The options are selected from a predefined list of CPU features based on the host architecture (Intel or AMD).


\subsection{Agent Program}
\label{anomaly_detection}

The agent program serves as the central coordinator of the overall fuzzing operation.  
In addition to managing communication between AFL++, the fuzz-harness VM, and the target L0 hypervisor, it performs the following roles:

\textbf{Fuzzing Orchestration:}  
The agent orchestrates the entire fuzzing loop for each test case.  
It first launches the UEFI executor in the fuzz-harness VM (using the \texttt{qemu-kvm} command for KVM), and then terminates it upon completion of the fuzzing iteration or detection of a potential vulnerability.  
During execution, the agent collects coverage data from the L0 hypervisor and passes it to AFL++ via shared memory after the VM terminates.

To decouple the fuzz-harness VM from the L0 hypervisor, the UEFI executor is designed to be self-contained.  
The agent embeds the fuzzing input as binary data into the UEFI binary in each execution, allowing the executor to run independently without interacting with the fuzzer during execution.  
This design improves portability and reduces dependency on specific execution environments.

\textbf{Vulnerability Detection:}  
The agent program also detects potential vulnerabilities and saves corresponding reports for reproduction.  
To identify anomalies that may indicate such vulnerabilities, it relies on hypervisor-specific bug detection mechanisms.  
For example, in KVM, the agent uses Kernel Address Sanitizer (KASAN) and Undefined Behavior Sanitizer (UBSAN), and monitors kernel log messages for relevant anomalies.  
In Xen, it monitors hypervisor-specific diagnostic logs for assertion failures, critical warnings, or other signs of unexpected hypervisor behavior.

Upon detecting an anomaly or observing new code coverage (indicated by an internal flag), the agent saves the current fuzzing input to a timestamped file within a designated directory specified in its configuration.  
This ensures that any crashes or unique behaviors can be reliably reproduced for subsequent manual analysis and debugging.



\section{Evaluation}

In this section, we evaluate \sysname through a series of experiments.  
The primary goal is to answer the following research questions:

\begin{itemize}
  \item \textbf{RQ1:} How much does \sysname improve code coverage compared to existing techniques?
  \item \textbf{RQ2:} How effective is each component of \sysname in improving code coverage?
  \item \textbf{RQ3:} Is \sysname effective across different hypervisors?
  \item \textbf{RQ4:} Is \sysname effective at finding vulnerabilities?
\end{itemize}

\subsection{Experimental Setup}
\label{sec:setup}

In KVM, we conducted fuzzing experiments using Linux kernel version 6.5 on Intel Core i9-12900K and AMD Ryzen Threadripper PRO 5995WX.  
Code coverage was measured with KCOV~\cite{kcov}, which collects instruction pointers for executed basic blocks via compile-time instrumentation.  
We mapped these pointers to source lines using \texttt{addr2line}.

To focus on nested virtualization-specific code that other techniques cannot reach, we restricted coverage measurement to \texttt{linux/arch/x86/kvm/\{vmx,svm\}/nested.c}.  
These files contain the core nested virtualization logic, including VMCS/VMCB emulation, consistency checks, and nested VM-exit handling.  
Other files also include code that supports nested virtualization, such as \texttt{vmx.c}/\texttt{svm.c} for general vCPU management, \texttt{mmu.c} for shadow paging, \texttt{tdp\_mmu.c} for two-dimensional paging (nested paging), and \texttt{posted\_intr.c} and \texttt{lapic.c} for interrupt processing.  
However, this code is intermingled with non-nested functionality, making its coverage less specific and more prone to noise.  
We confirmed that all L2-to-L0 and nested L1-to-L0 VM exits eventually dispatch to the handlers in \texttt{nested.c}, confirming that these files capture the code paths uniquely triggered by nested virtualization.  
Extending coverage measurement to related but less direct code paths remains an interesting direction for future work.  

For comparison, we evaluated two fuzzers: IRIS~\cite{10202630} and Syzkaller~\cite{syzkaller}.
IRIS is a state-of-the-art fuzzer for the Xen hypervisor that mutates VMCS data within Xen itself.
Although it does not target nested virtualization, we ran IRIS inside an L1 VM on L0 KVM to indirectly stress nested virtualization code.  
Syzkaller is the only available fuzzing tool that explicitly targets nested virtualization via manually written harnesses; we configured it to focus exclusively on KVM's \texttt{ioctl} interface to accelerate coverage.

We also considered other fuzzing frameworks.
However, HyperPill~\cite{298158} crashed when executed in a VM, ViDeZZo~\cite{10179354} does not use hardware-assisted virtualization, and HyperFuzzer~\cite{10.1145/3460120.3484748} is closed-source.  
Even when these tools run, they do not target nested virtualization interfaces and thus fail to improve coverage, as our IRIS experiments also suggest.

In addition, we used two testing tools: Selftests~\cite{linux_selftest} and KVM-unit-tests~\cite{kvm-unit-tests}.  
Selftests, included in the Linux kernel source tree, provides various KVM test programs.  
We executed all of them and aggregated their coverage, excluding overlaps.  
KVM-unit-tests is a minimal guest OS that implements unit tests for KVM.  
These tools directly exercise hardware-assisted virtualization code.

In Xen, we used Xen version 4.18 on Intel Core i9-12900K and AMD Ryzen 9 5950X.  
Code coverage was measured using gcov~\cite{gcov}.  
To focus on nested virtualization, we restricted the measurement to the following source files: \texttt{xen/arch/x86/hvm/\{vmx/vmx, svm/nestedsvm\}.c}.  
For comparison, we used the Xen Test Framework (XTF)~\cite{xen-xtf} as a baseline test suite.


In measuring coverage, we follow the guidelines of Klees et al.~\cite{10.1145/3243734.3243804}, reporting medians of five runs over time together with their 95\% confidence intervals (CIs), the $p$-values from two-sided Mann Whitney U-tests, and Cohen’s $d$ effect sizes.

\subsection{RQ1. Coverage Improvement}
\label{subsec:rq1}

\begin{table}[t]
  \centering 
  \begin{tabular}{l|r|r|r|r}
  \hline
   & \multicolumn{2}{c|}{Intel} & \multicolumn{2}{c}{AMD} \tabularnewline
   \hline
   & cov\% &\#line & cov\% &\#line \tabularnewline
  \hline \hline
  Total & 100\% & 1,681 & 100\% & 387 \tabularnewline
  \hline \hline
   \sysname & 84.7\% & 1,423 & 74.2\% & 287 \tabularnewline
  \hline
  Syzkaller & 61.4\% & 1,032 & 7.0\% & 27 \tabularnewline
  Syzkaller-\sysname & 7.3\% & 122 & 1.3\% & 5 \tabularnewline
  \sysname-Syzkaller & 30.5\% & 513 & 68.5\% & 265 \tabularnewline
  NecoFuzz$\cap$Syzkaller & 54.1\% &910 & 5.7\% & 22 \tabularnewline
  \hline
  IRIS & 52.3\% & 880 & - & - \tabularnewline
  \hline
  Selftests & 57.8\% & 971 & 73.4\% & 284 \tabularnewline
  Selftests-\sysname & 2.4\% & 41 & 8.0\% & 31 \tabularnewline
  \sysname-Selftests & 29.3\% & 493 & 8.8\% & 34 \tabularnewline
  NecoFuzz$\cap$Selftests & 55.3\% & 930 & 65.4\% & 253 \tabularnewline
  \hline
  KVM-unit-tests & 72.0\% & 1,211 & 69.8\% & 270 \tabularnewline
    \hline
\end{tabular}
\caption{KVM code coverage for nested virtualization-specific code.  
For \sysname and Syzkaller, the reported values are the median coverage after 48 hours of execution.  
For IRIS, the value corresponds to its final coverage upon termination.  
For Selftests and KVM-unit-tests, the values are measured after a single run.  
$A \cap B$ denotes the lines covered by both A and B, and $A - B$ denotes the lines covered by A but not by B.}
\label{tab:cov}
\end{table}

\begin{figure*}[t]
    \centering
  \begin{minipage}[t]{0.48\linewidth}
    \centering
    \includegraphics[width=\linewidth]{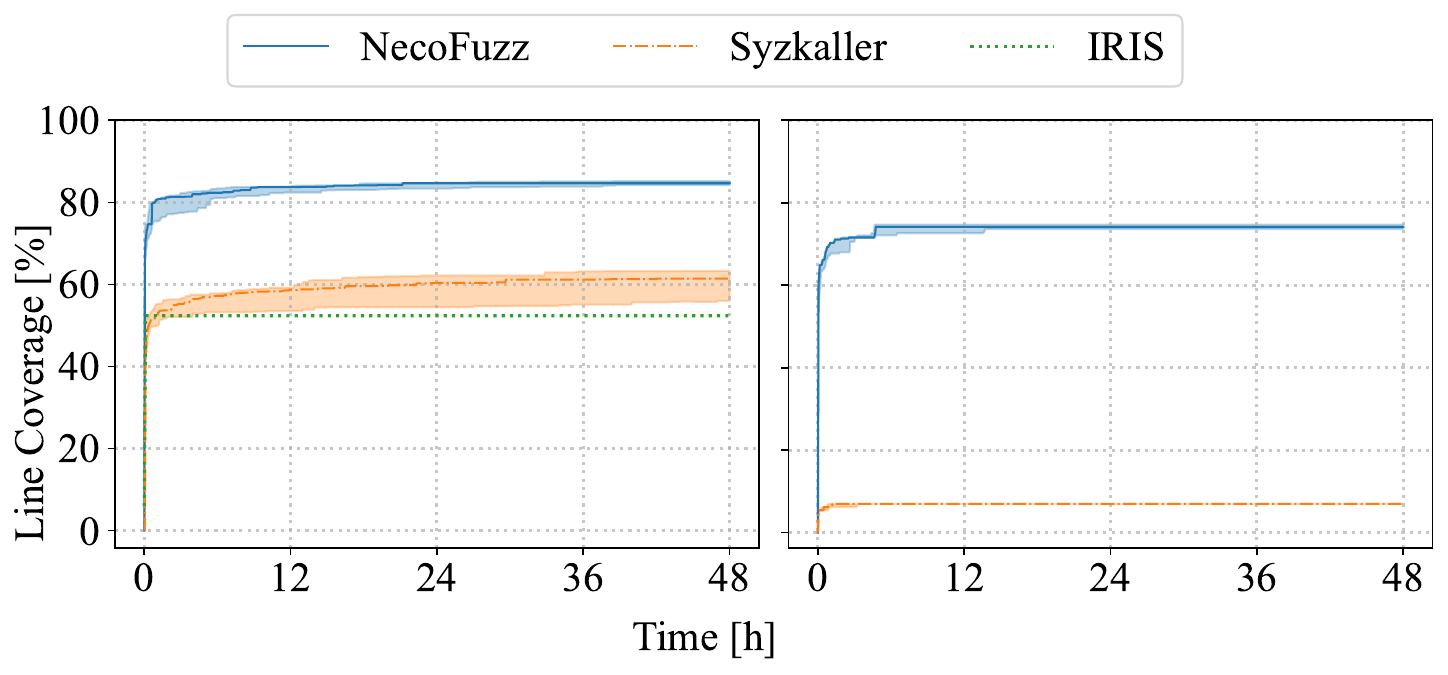}
    \begin{subfigure}{0.48\linewidth}
      \caption{Intel}
      \label{fig:cov:intel}
    \end{subfigure}
    \hfill
    \begin{subfigure}{0.48\linewidth}
      \caption{AMD}
      \label{fig:cov:amd}
    \end{subfigure}
    \captionsetup{width=0.9\linewidth}
    \caption{
      Code coverage transition over 48 hours for nested virtualization-specific code. (a) Intel, (b) AMD. IRIS indicates the maximum coverage at termination. Statistical significance: Intel ($p < 0.05$), AMD ($p < 0.05$).
    }
    \label{fig:cov:transition}
  \end{minipage}
  \hfill
  \begin{minipage}[t]{0.51\linewidth}
    \centering
    \includegraphics[width=1.\linewidth]{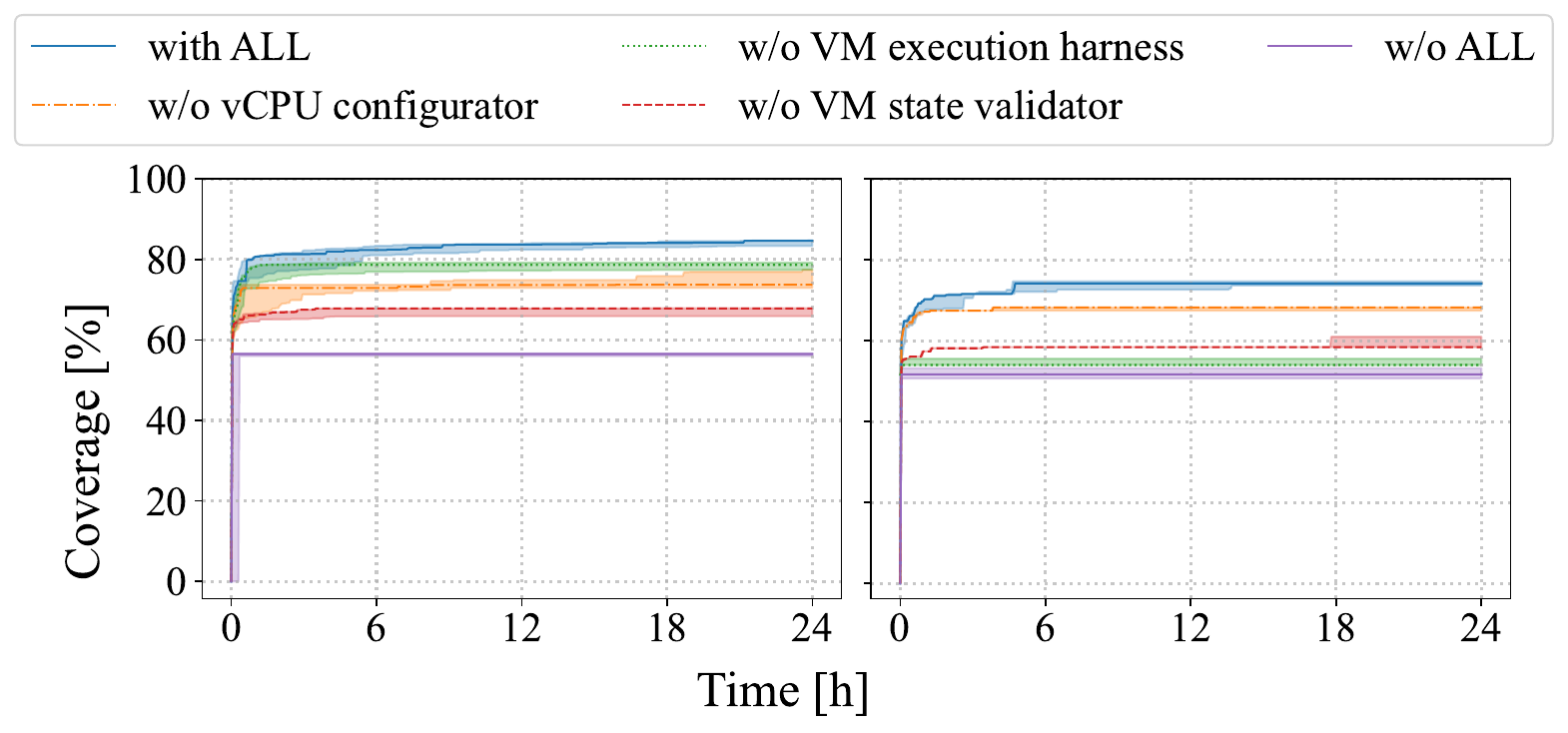}
    \begin{subfigure}[t]{0.48\linewidth}
    \vspace*{-4mm}
      \caption{Intel}
      \label{fig:cov:component:intel}
    \end{subfigure}
    \hfill
    \begin{subfigure}[t]{0.48\linewidth}
    \vspace*{-4mm}
      \caption{AMD}
      \label{fig:cov:component:amd}
    \end{subfigure}
    \captionsetup{width=0.8\linewidth}
    \caption{
      Breakdown of code coverage contribution for each VM generator component in \sysname. (a) Intel, (b) AMD.
    }
    \label{fig:cov:component}
  \end{minipage}
\end{figure*}

To answer RQ1, we measured the code coverage progression in KVM using \sysname and compared it with IRIS, Syzkaller, Selftests, and KVM-unit-tests.  
We ran them for 48 hours.  
IRIS was unstable in the nested environment and crashed after a few minutes; we report the coverage at the point of termination.  

\autoref{tab:cov} shows the measured results.
On Intel, \sysname achieved a median coverage of 84.7\% (95\% CI: 84.2--85.2), a 1.4$\times$ improvement over Syzkaller, which reached 61.4\% (95\% CI: 56.0--63.3), with $p = 0.012$ and a large effect size (Cohen's $d = 12.17$).  
On AMD, \sysname achieved 74.2\% (95\% CI: 73.6--74.7), corresponding to an 11.0$\times$ improvement over Syzkaller’s 7.0\% (95\% CI: 7.0--7.0), with $p = 0.014$ and Cohen's $d = 171.97$.
The code covered uniquely by Syzkaller (Syzkaller-\sysname) accounted for only 7.3\% on Intel and 1.3\% on AMD, whereas \sysname-uniquely covered code (\sysname-Syzkaller) reached 30.5\% on Intel and 68.5\%, respectively.
Note that Syzkaller lacks an AMD-specific harness, resulting in limited coverage of nested virtualization code.  
These results demonstrate that \sysname significantly improves code coverage while also subsuming nearly all lines reached by Syzkaller.

The uncovered code in our experiments can be classified into two categories.  
The first category consists of functions that can only be invoked by host-side operations, such as live migration, nested state setup and teardown, and hardware initialization or cleanup routines executed during KVM module load and unload.  
As stated in \autoref{subsec:threat_model}, functions that cannot be triggered directly by instruction execution in L1 or L2 guests are out of scope for our threat model.  
These functions are called via ioctl(), which accounts for approximately 4.8\% on Intel and 9.8\% on AMD, although precisely identifying which ioctl() operations are inaccessible from the guest is difficult.  

The second category consists of code executed only in rare situations, including bug-detection checks such as \texttt{BUG()}, memory-allocation failure handlers, support for minor hardware features such as Intel PT, Intel SGX, and posted-interrupt handling, and hypervisor-specific support such as enlightened VMCS for Hyper-V, which together account for up to 2.0\%.  
It also includes a small fraction of code that cannot be exercised with random seeds, such as instructions requiring specific operand values or VMCS configurations.  
Covering this residual code (in total less than 10\%) would require enhanced guest-OS fuzzing harnesses to initialize the corresponding execution environments and trigger the appropriate events.  

IRIS reached 52.3\% coverage almost immediately, which \sysname still outperforms by 1.6$\times$, and its coverage saturated quickly even within a few minutes.  
Although we could not run IRIS for an extended period, further significant improvement is unlikely, as its coverage of valid VMCS states is limited.
Compared to Selftests, which directly invoke nested virtualization code via \texttt{ioctl}, \sysname achieved 2.1$\times$ higher coverage on Intel and 1.1$\times$ on AMD.
Selftests-only coverage (``Selftests-\sysname'') was 2.4\% on Intel and 8.0\% on AMD, while \sysname-exclusive coverage (``\sysname-Selftests'') was 29.3\% on Intel and 8.8\% on AMD.
Compared to KVM-unit-tests, which consists of manually written unit tests, \sysname covered 1.18$\times$ more lines on Intel and 1.06$\times$ more on AMD.
Note that increasing coverage in unit tests can be easily achieved by manually writing test code that exercises the target code, but such tests do not necessarily explore complex arguments and therefore may not lead to higher vulnerability detection rates.
In fact, Selftests run only 60 test cases in about 80 seconds, and KVM-unit-tests run only 84 test cases in about 20 minutes, both with a fixed set of deterministic test cases, whereas our fuzzing framework stresses the target code for 48 hours or longer.

\autoref{fig:cov:intel} and \autoref{fig:cov:amd} show the coverage progression over 48 hours for \sysname and Syzkaller (IRIS is shown as a dotted horizontal line for reference; it crashed after a few minutes).  
Both begin with moderate coverage due to their respective harnesses.  
\sysname rapidly increases coverage---from approximately 70\% to 84.7\% on Intel, and from 65\% to 74.2\% on AMD.  
In contrast, Syzkaller exhibits much slower convergence: on Intel, it reached only 61.4\% after 48 hours. 
These results demonstrate that \sysname efficiently expands the code coverage of nested virtualization logic by mutating input VMs more effectively.


\subsection{RQ2. Component Effectiveness}

To answer RQ2, we evaluated the contribution of each component by selectively disabling it and measuring the resulting coverage.  
We also analyzed the distribution of generated VM states to assess their effectiveness.

\subsubsection{Breakdown of Coverage Contribution}

\begin{table}[t]
  \centering 
  \begin{tabular}{l|r|r}
  \hline
   & \multicolumn{1}{c|}{Intel} & \multicolumn{1}{c}{AMD} \tabularnewline
  \hline
  with ALL & 84.7\% & 74.2\% \tabularnewline
  w/o VM execution harness & 78.6\% & 54.0\% \tabularnewline
  w/o VM state validator & 67.8\% & 58.4\% \tabularnewline
  w/o vCPU configurator & 73.7\% & 68.2\% \tabularnewline
  w/o ALL & 56.5\% & 51.7\% \tabularnewline
  \hline
  \end{tabular}
\caption{Contribution of each component of \sysname, showing median coverage across five runs after 24 hours.}
\label{tab:component}
\end{table}

We measured the code coverage of \sysname with each component of the VM generator selectively disabled.  
\autoref{tab:component} presents the coverage at the 24-hour mark, while \autoref{fig:cov:component:intel} and \autoref{fig:cov:component:amd} show the coverage progression over time.  
Disabling any single component resulted in a noticeable decrease compared to the full configuration (“with ALL”).  
On Intel, removing the VM execution harness reduced coverage by 6.1 percentage points (pp), the VM state validator by 16.9 pp, and the vCPU configurator by 11.0 pp. 
On AMD, the respective reductions were 20.2 pp, 15.8 pp, and 6.0 pp.

Among the components, the VM state validator had the largest impact on Intel (16.9 pp), whereas the VM execution harness was most critical on AMD (20.2 pp).  
When all components were disabled (``w/o ALL''), only a predefined template and a default vCPU configuration were used, resulting in a coverage drop of 28.2 pp on Intel and 22.5 pp on AMD.
These results demonstrate that all three components make meaningful contributions to improving the coverage of nested virtualization code.


\subsubsection{Distribution of VM States}

\begin{figure}[tb]
  \centering
  \includegraphics[width=\columnwidth]{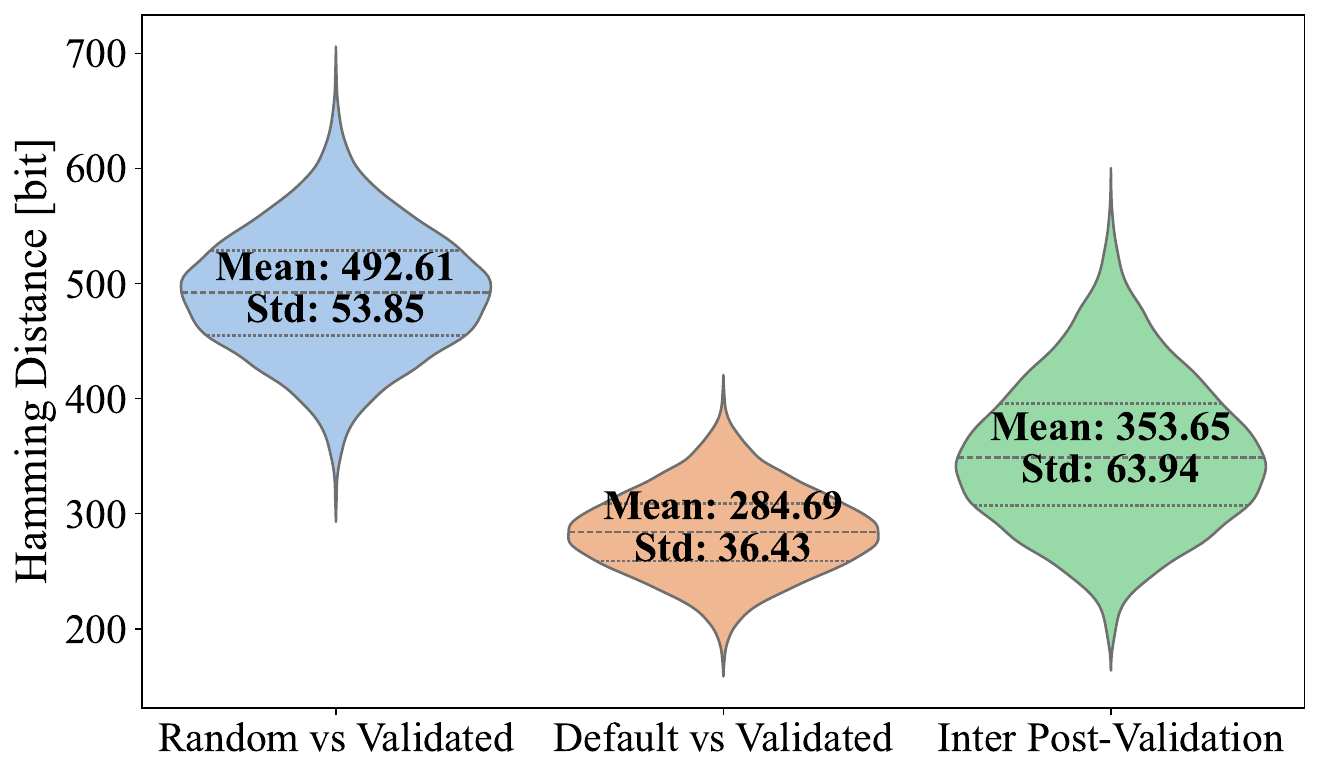}
  \caption{Distribution of VM states. The violin plot shows the Hamming distance distribution, with the mean and standard deviation indicated.}
  \label{fig:hamming}
\end{figure}

We measured the Hamming distance (i.e., number of differing bits) between randomly generated VM states and their validated counterparts produced by the VM state validator to evaluate its effectiveness.  
We used the VMCS layout, which defines an 8,000-bit VM state across 165 fields with predefined widths.  
This experiment was repeated 10{,}000 times. The result is visualized on the left side of \autoref{fig:hamming}.
The average Hamming distance was 492.6 bits with a standard deviation of 53.9, indicating that randomly generated VM states have an extremely low probability of matching valid ones, approximately one in $2^{492.6}$.

To evaluate state diversity, we compared generated VM states to those derived from simple default-initialized values.  
As shown in the center of \autoref{fig:hamming}, the average Hamming distance was 284.7 bits with a standard deviation of 36.4.  
This suggests that the validator produces more diverse states than can be achieved through simple default mutations.

Finally, to assess intra-set variability, we computed the Hamming distance between pairs of generated VM states.  
The right side of \autoref{fig:hamming} shows an average distance of 353 bits with a standard deviation of 63.9.  
These results confirm that the validator generates a wide variety of VM states that are both near-valid and internally diverse.

\subsection{RQ3. Hypervisor Independence}
\label{subsec:rq3}

\begin{table}[t]
  \centering 
  \begin{tabular}{l|r|r|r|r}
  \hline
   & \multicolumn{2}{c|}{Intel} & \multicolumn{2}{c}{AMD} \tabularnewline
   & cov\% &\#line & cov\% &\#line \tabularnewline
  \hline \hline
  Instrumented & 100\% & 1,401 & 100\% & 794 \tabularnewline
  \hline \hline
  \sysname & 83.4\% & 1,168 & 79.0\% & 627 \tabularnewline
  \hline
  XTF & 20.4\% & 286 & 10.8\% & 86 \tabularnewline
  NecoFuzz$\cap$XTF & 18.7\% & 262 & 8.4\% & 67 \tabularnewline
  \sysname-XTF & 64.7\% & 906 & 70.5\% & 560 \tabularnewline
  XTF-\sysname & 1.7\% & 24 & 2.4\% & 19 \tabularnewline
  \hline
\end{tabular}
\caption{Xen code coverage of nested virtualization-specific code after 24 hours, showing median coverage across five runs.}
\label{tab:xen}
\end{table}

To demonstrate the versatility of \sysname, we extended our evaluation to the Xen hypervisor.
\autoref{tab:xen} compares the code coverage achieved by \sysname and the XTF after 24 hours of fuzzing.
The results are consistent with the trends observed in \autoref{tab:cov} for KVM.
\sysname achieved 83.4\% coverage on Intel (95\% CI: 81.8--83.9) and 79.0\% on AMD (95\% CI: 77.7--80.5), outperforming XTF by 63.0 and 68.2 percentage points, respectively.
These results confirm that \sysname provides significantly higher coverage than existing approaches across different hypervisors.


\begin{table}[t]
  \centering 
  \begin{tabular}{l|l|l}
  \hline
   & \multicolumn{1}{c|}{Intel} & \multicolumn{1}{c}{AMD} \tabularnewline
  \hline
  w/o coverage guidance & 84.7\% & 74.2\%\tabularnewline
  with coverage guidance & 81.7\%  & 71.8\%\tabularnewline
  \hline
\end{tabular}
  \caption{The effect of coverage guidance in \sysname.}
  \label{tab:wo_cov}
\end{table}




Next, to evaluate the applicability of \sysname to closed-source hypervisors, we assessed the impact of coverage guidance in KVM.  
\autoref{tab:wo_cov} shows the results after 48 hours of fuzzing.  
Interestingly, enabling or disabling coverage guidance had only a minor effect on the final coverage.  
This may be because, particularly when exploring the VMCS state space, broadly testing subtle boundary regions using the VM state validator is more effective for improving coverage than following deep execution paths.  
These results suggest that \sysname can serve as an effective black-box fuzzer for closed-source hypervisors that do not expose coverage information.

\subsection{RQ4. Vulnerability Detection Capabilities}

\begin{table*}[ht]
\centering
\begin{tabular}{|c|c|c|c|c|c|}
\hline
No & Hypervisor & CPU & Cause & Detection Method & Status \\
\hline
1 & KVM & Intel & VM State Handling Flaw & UBSAN & Fixed~\cite{patch-kvm-cve}, CVE-2023-30456~\cite{CVE-2023-30456} \\
2 & VirtualBox & Intel & VM State Handling Flaw & VM Crash & Fixed, CVE-2024-21106~\cite{CVE-2024-21106} \\
3 & KVM & Intel, AMD & Page Table Handling Flaw & Assertion & Fixed~\cite{patch-KVM-bug} \\
4 & Xen & Intel & VM State Handling Flaw & Host Crash & Fixed~\cite{patch-xen} \\
5 & Xen & AMD & VM State Handling Flaw & Assertion & Confirmed~\cite{xen-issue-215} \\
6 & Xen & AMD & VM State Handling Flaw & Assertion & Confirmed~\cite{xen-issue-216} \\
\hline
\end{tabular}
\caption{Newly discovered vulnerabilities in nested virtualization across different hypervisors.}
\label{tab:hypervisor_vulnerabilities}
\end{table*}

To evaluate the vulnerability detection capabilities of \sysname, we examined whether it could uncover real-world bugs in existing hypervisors.  
We applied \sysname to KVM, Xen, and VirtualBox, and analyzed fuzzer outputs and execution traces to confirm issues.  
As a result, \sysname identified six previously unknown vulnerabilities: two in KVM, three in Xen, and one in VirtualBox.  
All were reported to the respective developers, and four have since been confirmed and fixed.  
Two of the confirmed vulnerabilities have been assigned CVEs—one in KVM and one in VirtualBox.  
\autoref{tab:hypervisor_vulnerabilities} summarizes the discovered vulnerabilities.  
Below, we describe selected examples in more detail.

\subsubsection{New Vulnerabilities in KVM}

We first describe how \sysname discovered CVE-2023-30456~\cite{CVE-2023-30456}, a bug in KVM's nested virtualization implementation on Intel.  
The issue stems from a missing consistency check in VMCS12, which leads to an array-index-out-of-bounds write error during L2 guest page walks by the L1 hypervisor.

To trigger this bug, the KVM module must be loaded with nested virtualization enabled but EPT disabled.  
The L1 hypervisor then sets the “IA-32e mode guest” bit in the VM-entry control field to 1, while leaving the GUEST CR4.PAE bit unset (0).  
Although the Intel Software Developer Manuals~\cite{intel-sdm} states that CR4.PAE must be set when IA-32e mode is enabled, the CPU silently assumes it is set and allows the VM entry to proceed.
KVM, however, interprets CR4.PAE literally and mismanages page tables, resulting in corruption.  
Thanks to the VM state validator and vCPU configurator, \sysname was able to produce this edge case automatically.

\sysname also discovered a second bug in KVM's handling of shadow roots during nested EPT operations.  
Here, the L1 hypervisor assigns an invalid EPT pointer (EPTP) in VMCS12 and launches the L2 guest.  
KVM attempts to validate the EPT root via \texttt{mmu\_check\_root()}, and upon failure, it incorrectly triggers a VM exit due to a triple fault---even though the L2 VM never actually starts.  
This erroneous behavior was exposed by VM states generated through the state validator.
To address this, the KVM developers applied a patch~\cite{patch-KVM-bug} that loads a dummy root from the zero page.  
This ensures that any guest memory accesses after L2 entry generate a clean fault and trigger a correct VM exit, avoiding incorrect transitions from L2 to L1.

\subsubsection{New Vulnerabilities in Xen}

\sysname uncovered three new bugs in Xen’s nested virtualization implementation.
The first bug occurs when nested virtualization is enabled in a hardware virtual machine (HVM) guest, and the L1 hypervisor sets the activity state field of VMCS12 to 3 (``wait-for-SIPI'') before executing \texttt{vmlaunch} to L2.  
This causes not only the guest VM to hang, but also the entire host system to become unresponsive.  
The issue stems from Xen’s nested virtualization logic blindly copying the activity state from VMCS12 into VMCS02.  
Some activity states, such as SHUTDOWN or WAIT-FOR-SIPI, are intended for auxiliary processor management in Intel Trusted Execution Technology (TXT) and block or filter events differently than the normal active state.  
For example, setting the activity state to SHUTDOWN triggers a platform reset, while WAIT-FOR-SIPI blocks all interrupts except SIPIs.  
These states are not meant to be used in normal nested VM configurations.  
Xen’s failure to sanitize the activity state during VM entry to L2 results in undefined behavior~\cite{patch-xen}.

The second and third bugs appear on AMD platforms, which use the Virtual Machine Control Block (VMCB) instead of Intel’s VMCS.  
The second bug occurs when the L1 hypervisor sets CR0.PG = 0 in VMCB12 after previously running a 64-bit L2 guest.  
This creates an inconsistent state: the Long Mode Enable (LME) bit in the Extended Feature Enable Register (EFER) is set to 1, while CR0.PG (paging) is cleared.  
Normally, paging must be enabled for long mode to function, so this combination is invalid.  
The inconsistency leads to memory corruption and causes the Advanced Virtual Interrupt Controller (AVIC) to become erroneously enabled in VMCB02.  
As a result, Xen triggers a VM exit with reason \texttt{AVIC\_NOACCEL}, even though AVIC is not supported.  
This issue highlights an architectural ambiguity in AMD’s specification~\cite{amd-manual}, which permits such a VMCB state but does not clarify how \texttt{vmrun} should behave—posing difficulties for correct nested virtualization support.

The third bug is also caused by an invalid value in CR4 within VMCB12, set by the L1 hypervisor.  
Xen correctly fails the \texttt{vmrun}, causing a VM exit back to L1.  
However, during the VM exit emulation, the function \texttt{nsvm\_vcpu\_vmexit\_inject()} checks the Virtual GIF Enable (VGIF) status.  
If VGIF is enabled, it assumes that the virtual GIF (Global Interrupt Flag) is set.  
In the observed case, \texttt{vgif} was unexpectedly zero, leading to an assertion failure.  
Although this does not crash the system, it reveals a flaw in Xen’s assumption about default interrupt masking behavior when VGIF is enabled.

All three bugs were uncovered using VM states generated by \sysname’s VM state validator, demonstrating its ability to exercise edge cases across architectures.

\subsubsection{New Vulnerabilities in VirtualBox}

Finally, \sysname discovered CVE-2024-21106~\cite{CVE-2024-21106}, a vulnerability in Oracle VirtualBox version 7.0.12 related to improper validation of Model-Specific Register (MSR) load values during nested VM entry.  
This issue allows non-canonical addresses to be written to MSRs, resulting in a general protection fault in the host.

To reproduce this bug, VirtualBox must run with nested virtualization.  
The L1 hypervisor sets a non-canonical address (e.g., 0x8000000000000000) in the \texttt{vmentry\_msr\_load} field of VMCS12 for MSR \texttt{KernelGSBase} (0xC0000102), then executes \texttt{vmlaunch}.  
According to Intel’s specification~\cite{intel-sdm}, addresses loaded into MSRs such as \texttt{KernelGSBase} must be canonical.  
However, VirtualBox fails to validate this condition during nested VM entry.

The vulnerability manifests in two ways:  
(1) it causes the VM to terminate unexpectedly, and  
(2) it prevents the VM from shutting down properly, leaving the system frozen.  
System logs report the following error:  
{"general protection fault, probably for non-canonical address 0x8000000000000000"}.

This behavior contrasts with KVM, which properly validates MSR values during nested entry.  
For example, KVM explicitly checks whether loaded MSR values are canonical and aborts the entry if not.  
The failure in VirtualBox thus reflects a gap in its validation logic.

\sysname triggered this bug by generating a boundary-state VMCS with a non-canonical MSR value using its VM state validator, highlighting the effectiveness of our specification-guided approach.

\subsection{Lessons learned on input generation}

We learned that combining rounding to a valid VMCS and then selectively injecting invalid values is highly effective for generating useful fuzzing inputs. Rounding moves raw mutations into the vicinity of valid states, which prevents early rejection and enables the fuzzer to exercise validation logic.
Selective injections of invalid bits after rounding then push the VMCS across subtle validity boundaries, increasing the likelihood of exposing error-prone handling code.
We also observed that traditional coverage guidance contributed less than expected in this setting.
Because rounding normalizes many small mutations, inputs suggested by coverage feedback often collapse to equivalent post-rounding states and thus provide limited additional signal.
Consequently, a breadth-first strategy that explores a wide range of near-valid states produced better coverage than a depth-first strategy that follows coverage-guided micro-variations.
Based on these observations, we suggest prioritizing state-space diversity around specification boundaries and combining lightweight validation-aware transformations with targeted invalidation as a practical input generation recipe for nested virtualization fuzzing.

\section{Discussion}

In this section, we discuss the potential limitations in bug detection and the narrow scope of the targeted code.



\subsection{Bug Detection}

In \sysname, we used existing sanitizers, such as UBSAN and KASAN, included in KVM.
However, we were unable to detect some known CVEs~\cite{CVE-2019-3887,CVE-2021-3656,CVE-2021-29657} despite attempts to reproduce and identify them.  
While bug detection is a critical aspect of fuzzing, vulnerabilities in nested virtualization often require specialized detection mechanisms, such as identifying misuse of L0 hypervisor privileges from within an L1 VM, rather than simply detecting crashes in the L0 hypervisor.  
This highlights the limitation of relying solely on sanitizers for comprehensive vulnerability detection and underscores the need for more advanced bug detection techniques tailored to the semantics of nested virtualization.

\subsection{Code Coverage}

\sysname focuses exclusively on measuring the coverage of code specific to nested virtualization.  
This focus was chosen because most existing hypervisor fuzzers provide no coverage for these parts of the code, making it essential to first evaluate the reachability of this region for effective nested virtualization fuzzing.  
However, other hypervisor subsystems, such as memory management and interrupt handling, may be indirectly related to nested virtualization and could contain relevant vulnerabilities.  
Identifying such code is challenging, as it is spread across multiple source files and is not explicitly labeled as nested virtualization logic.  
While further improving coverage of nested virtualization-specific code remains important, future work should investigate designing specialized harnesses that can also target code paths indirectly connected to nested virtualization.

\subsection{Asynchronous Events}

\sysname focuses on VM exits explicitly triggered by guest instructions.
These VM exits are deterministic and easier to control via fuzzing inputs.
We do not currently model asynchronous external events (e.g., interrupts, NMIs, timer-based VM exits), as they require precise event injection and temporal control, which complicate repeatability and determinism.
While such events are important for full system testing, prior studies~\cite{10.1145/3190508.3190529,10.1145/3460120.3484748} show that instruction-triggered VM exits cover the majority of security-relevant hypervisor logic.
We leave modeling and injection of asynchronous events to future work.

\subsection{Limitations}

\sysname currently implements a single-vCPU guest harness, which limits its ability to find bugs arising from interactions across multiple CPUs.  
In addition, \sysname supports only a single nested VM and does not cover bugs caused by inter-VM communication.
Deeper nesting, such as supporting an L3 guest OS, is also not supported.
To explore these types of inter-domain interactions (e.g., inter-vCPU, inter-VM, and inter-layer), fuzz-harness VMs would need to implement corresponding communication mechanisms and coordinate with the fuzzer to exercise them.
Although these communication paths are not necessarily directly tied to the nested virtualization logic in the host hypervisor, they may still contain code fragments that are executed only in nested environments, similar to other cases discussed in \autoref{sec:setup}.  
We consider extending \sysname to support such communication an interesting direction for future work.

\section{Related Work}

\sysname is the first fuzzing framework that effectively targets nested virtualization-specific code.  
To the best of our knowledge, NestFuzz~\cite{nestfuzz} is the only prior work that explicitly attempts to fuzz nested virtualization.  
However, it is an early-stage work that issues random VMX instructions without addressing key challenges such as VM state validity, initialization sequences, or execution harnessing, and it lacks evaluation of code coverage or vulnerability detection.

Prior research on hypervisor fuzzing can be broadly classified into two categories based on their primary targets: \emph{virtual CPU fuzzing} and \emph{virtual device fuzzing}.

\subsection{Virtual CPU Fuzzing}

Prior work on virtual CPU fuzzing has not addressed the emulation of hardware-assisted virtualization.
While some studies attempt to bring target CPUs into initialized states using templates, they overlook the challenges posed by the vast VM state space and the complex interdependencies in vCPU configuration during VM startup.

We further categorize these efforts into two types: \emph{blackbox fuzzing}, which treats the target as opaque and does not require internal knowledge, and \emph{non-blackbox fuzzing}, which leverages internal instrumentation or symbolic reasoning.

\paragraph{Blackbox Fuzzing.}

Blackbox fuzzing operates without source code or internal knowledge of the target hypervisor.
It enables fast deployment but often yields lower coverage due to limited semantic insight.

EmuFuzzer~\cite{10.1145/1572272.1572303} applies fuzzing to CPU emulators to detect behavioral deviations from real CPUs.
It infers instruction formats empirically on real hardware, allowing broad instruction set coverage without disassemblers.
However, it only supports user-mode instructions and cannot test privileged behavior relevant to hypervisors.
KEmuFuzzer~\cite{10.1145/1831708.1831730} extends EmuFuzzer to support kernel-mode instructions.
It uses templates with symbolic fields to generate a minimal guest OS that runs test cases.
While this produces some valid states, their diversity is limited.
It also lacks support for hardware-assisted virtualization initialization and cannot directly mutate VMCS-like structures.

Amit et al.~\cite{10.1145/2815400.2815420} adapt Intel CPU validation tools to test hypervisors by comparing results against Intel’s simulator.
Their test cases combine manual templates and random instruction sequences, with CPU settings specified via a configuration file.
While the paper mentions nested virtualization, it does not evaluate it due to KVM limitations and does not address its distinct challenges.

\paragraph{Non-blackbox Fuzzing.}

Non-blackbox fuzzing, including graybox, whitebox, and hybrid approaches, uses internal knowledge of the target hypervisor to guide test generation through coverage feedback or symbolic execution.
These techniques often require source access and are difficult to apply to closed-source hypervisors.

PokeEMU~\cite{10.1145/2150976.2151012} uses high-fidelity emulators such as Bochs to symbolically generate test cases for low-fidelity targets like QEMU.
It requires manual annotation of symbolic registers and memory, making it difficult to apply to complex structures like VMCS.

MultiNyx~\cite{10.1145/3190508.3190529} runs the hypervisor on top of an executable specification to symbolically explore its behavior.
Simple instructions execute directly, while complex ones, such as virtualization operations, are handled by the specification.
However, it does not validate VM state correctness and requires source code modification to annotate inputs.

HyperFuzzer~\cite{10.1145/3460120.3484748} combines coverage guidance with symbolic execution using Intel PT tracing to perform lightweight analysis on real hardware.
It focuses on control flow only, lacks VM state validation, and is limited to Intel PT-compatible systems.

IRIS~\cite{10202630} uses record-and-replay by collecting traces from guest OS executions and reusing them as seeds.
Since these come from well-behaved OSes, VM state diversity is limited, reducing effectiveness in exposing vulnerabilities.

Syzkaller~\cite{syzkaller}, a kernel fuzzer, supports KVM’s nested virtualization by invoking system calls and assigning random values to VM states.
This limits the variety of valid instruction sequences and reduces coverage.

Symbolic and concolic testing may appear effective for nested virtualization code bases, which involve complex instruction emulation yet remain relatively small in size.  
However, the main challenge in our setting arises from the vast VMCS state space with more than 150 interdependent fields; treating the entire structure as symbolic would overwhelm SMT solvers.  
Concolic testing further requires tight integration with the target, which is infeasible for closed-source hypervisors, while whitebox fuzzing demands accurate hardware models that are difficult to construct due to complex and partly undocumented CPU behavior.  
Nevertheless, lightweight symbolic or hybrid approaches could complement our fuzzing framework and remain an interesting direction for future work.

\subsection{Virtual Device Fuzzing}

Virtual device fuzzing targets I/O interfaces and has uncovered vulnerabilities in hypervisors like QEMU/KVM, VirtualBox, and VMware Fusion.

VDF~\cite{10.1007/978-3-319-66332-6_1} pioneered this direction by logging MMIO activity and seeding AFL-based fuzzing.
HYPER-CUBE~\cite{hyper-cube} performs black-box fuzzing on PIO/MMIO using a custom OS and bytecode interpreter, achieving high throughput without coverage guidance.
NYX~\cite{263866} builds structured fuzzing from user-defined protocol specifications, though it requires manual effort.
V-Shuttle~\cite{10.1145/3460120.3484811} intercepts DMA accesses and injects fuzzing data into nested DMA structures.
Morphuzz~\cite{277148} infers I/O ranges from guest address maps to improve fuzzing efficiency.
ViDeZZo~\cite{10179354} focuses on intra- and inter-message dependencies of virtual devices.
HyperPill~\cite{298158} uses hardware virtualization for snapshot-based fuzzing of QEMU devices.

However, these approaches do not apply to nested virtualization, which relies on hardware-assisted virtualization instructions, not I/O interfaces.
As a result, virtual device fuzzing fails to reach nested virtualization-specific code.

\section{Conclusion}

We presented \sysname, the first fuzzing framework designed specifically for nested virtualization in hypervisors.  
\sysname introduces a VM generator that efficiently synthesizes complete VM instances, called \emph{fuzz-harness VMs}, to explore security-critical logic near the boundary between valid and invalid VM states.  
The generator comprises three components: a VM execution harness that triggers nested virtualization code paths, a VM state validator that guides input generation using hardware specification models, and a vCPU configurator that systematically explores combinations of hardware-assisted virtualization features.
We implemented \sysname on both Intel VT-x and AMD-V and applied it to fuzz KVM, Xen, and VirtualBox.  

Our evaluation demonstrated that \sysname significantly improves coverage of nested virtualization-specific code compared to existing fuzzers, including Syzkaller, the only prior tool that explicitly supports nested virtualization.  
Furthermore, \sysname uncovered six previously unknown vulnerabilities across the three hypervisors, including two that have been assigned CVEs.
These results show that \sysname provides a practical and effective foundation for fuzzing the previously unexplored attack surface of nested virtualization.

\begin{acks}
This work was supported by JST, CREST Grant Number JPMJCR22M3, Japan. We thank our shepherd, Prof.\ Trent Jaeger, and the anonymous reviewers for their insightful feedback and guidance, as well as the anonymous artifact evaluation reviewers for their efforts and valuable comments toward reproducibility. We used DeepL and ChatGPT to assist with English translation and language refinement.
\end{acks}

\bibliographystyle{\Bibstyle}
\balance
\bibliography{main}
\end{document}